\documentclass[12pt]{article}
\usepackage{amsmath, amssymb}

\def\hybrid{\topmargin -20pt    \oddsidemargin 0pt
        \headheight 0pt \headsep 0pt
        \textwidth 6.25in       
        \textheight 9.5in       
        \marginparwidth .875in
        \parskip 5pt plus 1pt   \jot = 1.5ex}

\hybrid

\catcode`\@=11

\def\marginnote#1{}		
\begin{document}
\begin{titlepage}
\begin{center}
{\huge \sf {Mean Curvature Flow on Ricci Solitons}}
\vskip 0.9in
{\large \bf Efstratios Tsatis}
\vskip 0.2in
{\em Department of Physics, University of Patras \\
GR-26500 Patras, Greece\\
\footnotesize{\tt efstratiostsatis@hotmail.com}}\\
\end{center}

\vskip 1.0in
\centerline{\bf Abstract}
We study monotonic quantities in the context of combined geometric flows. In particular, focusing on Ricci solitons as the ambient space, we consider solutions of the heat type equation integrated over embedded submanifolds evolving by mean curvature flow and we study their monotonicity properties. This is part of an ongoing project with Magni and Mantegazzawhich will treat the case of non-solitonic backgrounds $\cite{a_14}$.
\end{titlepage}

\newpage
\tableofcontents\thispagestyle{empty}
\newpage

\numberwithin{equation}{section}
\section{Introduction}
\setcounter{equation}{0}
The subject of the geometric quantities evolving with respect to some time variable $t$ has been an arena of intense study in the framework of Riemannian geometry. The prime example in this framework has been the recent celebrated work of Perelman $\cite{perel_1}$, which has led to the verification of Thurston's Geometrization conjecture (see $\cite{a_2}$ for an excellent review). In general, there are two categories of geometric quantities and therefore two kinds of flows: \emph{intrinsic} quantities, such as the metric of the manifold, and \emph{extrinsic} quantities such as the embedding of a manifold in some ambient space. To the first category belong the Ricci flow, the Calabi flow and the Yamabe flow. To the second category belongs the mean curvature flow (MCF). The Ricci flow, being the prominent member of intrinsic flows, had been introduced by Hamilton in an effort to find metrics of constant Ricci curvature. The importance of the MCF lies in the fact that its fixed points are submanifolds of minimal volume embedded in some fixed space. An important attribute of the aforementioned flows is the fact that they are gradient. This means that they arise as equations of motion by extremizing some ``energy" functional.
\par
In the realm of Physics, the applications of both the Ricci flow and the MCF can not be underestimated. The former was first introduced by Friedan $\cite{a_3}$ and the latter by Leigh $\cite{a_4}$. The Ricci flow naturally describes the response of the generalized coupling constant $g_{i j}(x,t)$, which is the metric of the target space in the $\sigma$-model description, to the change of the energy scale $\Lambda$, defined as the exponent of $t$, to the lowest order in perturbation theory respectively. Analogously, the MCF describes the change of the geometry of defects (branes) embedded in some ambient space with respect to the energy scale. These two equations are the \emph{Renormalization Group Equations} (RGE) for the metric and the embedding respectively. The introduction of entropy-like functionals, which in most cases, are quantities of strict monotonicity along the flow, shed light in the concept of the $c-theorem$ introduced by Zamolodchikov $\cite{zamo_1}$. Recently, an attempt was made $\cite{a_5}$ to go beyond the first order in $\alpha'$, where Ricci flow and RGE flow coincide, to determine whether the monotonicity to all orders in $\alpha'$ holds. In addition to the case of the Ricci flow, quantities exhibiting monotonicity have been found in the case of MCF. In this case we can have submanifolds embedded in some fixed ambient space evolving according to their mean curvature vector. A monotonic quantity along the mean curvature flow has been found by Huisken and then generalized by Hamilton. In this context, the monotonic quantities are based on solutions to the heat equation.
\par
The purpose of this work is to consider the case of MCF evolution when the ambient space is also evolving by the Ricci flow, in an effort to find a quantity that remains monotonic. It is organized as follows: In section 2 we review some basic features of the Ricci flow focusing on monotonic quantities introduced by Perelman, such as the energy $\cal{F}$, the entropy $\cal{W}$ and the reduced length. In section 3 we turn our attention to solutions of the conjugate heat equation of the ambient manifold, in the case where an embedded submanifold evolves under mean curvature flow, while keeping the ambient space fixed. This setup serves as the basis, when we generalize the previous construction in section 4 to consider a Ricci soliton as the ambient manifold. We conclude with some directions for further research.

\section{Monotonic quantities for the Ricci flow}
\setcounter{equation}{0}
In this section, we set the stage by recalling certain facts pertaining to the evolution of metrics under the Ricci flow as well as we discuss some of the relevant advances made by Perelman towards the proof of the Poincare conjecture. In the first subsection, we will be reviewing certain monotone quantities (each of which is constant on the appropriate Ricci soliton) along the flow and comment on their physical interpretation. The second subsection will be devoted to the study of the reduced length. We will verify the fact that the potential $f$ (defined below) serves as a lower bound of the reduced lenght $\ell$ for the case of the cigar soliton.

\subsection{Ricci solitons and monotone functionals}

The Ricci flow assumes the form
\begin{equation}
\frac{\partial{g_{\mu\nu}}}{\partial{t}}=-2 R_{\mu\nu}
\end{equation}
where the term $R_{\mu\nu}$ is the Ricci curvature of the manifold. We are going to focus our attention on \emph{self-similar} solutions under the Ricci flow called solitons. These are solutions which evolve either by rescalings or by diffeomorphisms or both of the initial metric. One very important property of a soliton solution is that it saturates some sort of a lower bound of an energy functional. We will expound on this in the sequel.
\par
The most general form of a metric evolving under by both rescalings and diffeomorphisms has the form

\begin{equation}
g(t)=\sigma(t)\psi _{t}(g_{0})
\end{equation}
For the case of solitons it is sufficient to take $\sigma(t)$, the time-dependent rescaling, to be \emph{linear} (take $\sigma '(t)=-2\lambda$), $\psi _{t}$ is a time-dependent diffeomorphism and $g_{0}$ is the initial metric ($t=0$). Substituting this ansatz in the Ricci flow equation, we easily obtain

\begin{equation}
2 R_{\mu\nu}=2\lambda g_{\mu\nu}-g_{\mu\rho}\nabla _{\nu}V^{\rho}-g_{\rho\nu}\nabla _{\mu}V^{\rho}
\end{equation}
The metric which evolves under the above equation is called a \emph{Ricci soliton}. Depending on the sign of $\lambda$ we can have an expanding, a steady and a shrinking Ricci soliton as shown in the table 1. When $V^{\mu}=g^{\mu\nu}\nabla _{\nu}f$ the solution is called \emph{gradient soliton}. We note that the above form of the Ricci tensor ``drives" the metric by a diffeomorphism. In other words, all the metrics, from the starting one until the last one, belong to the same conjugacy class of the space of all the metrics modulo diffeomorphisms. A \emph{breather} is a metric which for some $t_{1}<t_{2}$ the metrics $a g_{\mu\nu}(t)$ and $g_{\mu\nu}(t)$ differ by a diffeomorphism. As a result, a soliton satisfies the breather condition at \emph{any} time interval. We have also the result that on a compact manifold, the notion of the breather and the gradient soliton coincide (for the case of a shrinking or a gradient soliton this was proved by Perelman, for the expanding soliton see $\cite{a_15}$)


\begin{table}[Bottom]\centering
\begin{tabular}{|l|l|}
\cline{1-2}
\vbox to1,88ex{\vspace{1pt}\vfil\hbox to12,80ex{\hfil $\lambda >0$ \hfil}} &
\vbox to1,88ex{\vspace{1pt}\vfil\hbox to21,20ex{\hfil     shrinking soliton    \hfil}} \\

\cline{1-2}
\vbox to1,88ex{\vspace{1pt}\vfil\hbox to12,80ex{\hfil $\lambda <0$ \hfil}} &
\vbox to1,88ex{\vspace{1pt}\vfil\hbox to21,20ex{\hfil     expanding soliton \hfil}} \\

\cline{1-2}
\vbox to1,88ex{\vspace{1pt}\vfil\hbox to12,80ex{\hfil $\lambda= 0 $\hfil}} &
\vbox to1,88ex{\vspace{1pt}\vfil\hbox to21,20ex{\hfil     steady soliton \hfil}} \\

\cline{1-2}
\vbox to1,88ex{\vspace{1pt}\vfil\hbox to12,80ex{\hfil $    V=0    $\hfil}} &
\vbox to1,88ex{\vspace{1pt}\vfil\hbox to21,20ex{\hfil     Einstein metric\hfil}} \\

\cline{1-2}
\end{tabular}
\caption{Ricci Solitons}
\end{table}

We proceed by recalling the definition and simple facts about the functionals $\cal{F}$ and $\cal{W}$ and their variation along the Ricci flow. They are defined over Riemannian metrics and smooth functions on $\cal{M}$. Let us start off with the functional $\cal{F}$. It assumes the form
\begin{equation}
{\cal{F}}(g ,f)=\int _{\cal{M}} \left(R+\left|\nabla f\right|^{2}\right)e^{-f} dV
\end{equation}
Now, suppose that the metric $g_{\mu\nu}$ and the function $f$ flow according to

\begin{equation}
\frac{\partial{g_{\mu\nu}}}{\partial{t}}=-2R_{\mu\nu},
\end{equation}
\begin{equation}
\frac{\partial {f}}{\partial{t}}=-\triangle _{\cal{M}}f-R+\left|\nabla f\right|^{2}
\end{equation}
under the condition that $\int _{\cal{M}}e^{-f}dV=1$ (this condition preserved by (2.5) and (2.6)). Then the time derivative of $\cal{F}$ along the flow is

\begin{equation}
\frac{d}{dt}{\cal{F}}\left(g_{\mu\nu},f\right)=2\int _{\cal{M}}\left|R_{\mu\nu}+\nabla _{\mu}\nabla _{\nu}f\right|^2 e^{-f}dV
\end{equation}
We observe that $\cal{F}$ is increasing along the flow unless we are in a steady gradient soliton where $\cal{F}$ is constant. Let us assume that the manifold $\cal{M}$ is compact without boundary and set $u=e^{-\frac{f}{2}}>0$. Integrating by parts the functional $\cal{F}$, we obtain,

\begin{equation}
{\cal{F}} \left(g,u\right)=\int _{\cal{M}} u\left(R-4\Delta\right)u\hspace {0.05in} dV
\end{equation}
We note that the operator ${\cal{A}}=R-4\Delta$ acts on ``wavefunctions" which by definition are strictly positive. Although quite technical (see $\cite{a_9}$ PART I p. 204), it can be proved that the lowest one is unique and it corresponds to the value of this functional. The aforementioned type of soliton is of great importance in string theory since its existence implies conformal invariance of the $\sigma$-model action at the quantum level (see $\cite{a_6,a_7}$ for related discussion on the two-dimensional Sausage model). In other words, the condition $R_{\mu\nu}+\nabla _{\mu}\nabla _{\nu}f=0$ is equivalent to the tracelessness of the energy-momentum tensor. As such it qualifies to be a background in string theory. In addition to this, the functional $\cal{F}$ is constant along the flow and it represents the correction of order $\alpha'$ to the central charge. Let us compute the value of the functional $\cal{F}$ for the case of a \emph{closed} steady soliton. We have $R=-\triangle _{\cal{M}}f$, therefore

\begin{equation}
{\cal{F}}\left(g,f\right)=\int _{\cal{M}} \left(-\triangle _{\cal{M}}f+\left|\nabla f\right|^{2}\right)e^{-f} dV=\int _{\cal{M}}\triangle _{\cal{M}}e^{-f} dV=0
\end{equation}

\par
Similar considerations can be made for the functional $\cal{W}$. Denoting $dim {\cal{M}}=n$, it is defined as,

\begin{equation}
{\cal{W}} \left(g,f,\tau\right)=\int _{\cal{M}} \left[\tau(R+|\nabla f|^{2})+f-n\right](4\pi\tau)^{-\frac{n}{2}}e^{-f} dV
\end{equation}
Now, suppose that the metric $g_{\mu\nu}$ and the function $f$ flow according to

\begin{equation}
\frac{\partial{g_{\mu\nu}}}{\partial{t}}=-2R_{\mu\nu}
\end{equation}
\begin{equation}
\frac{\partial {f}}{\partial{t}}=-\triangle _{\cal{M}}f-R+|\nabla f|^{2}+\frac{n}{2\tau}
\end{equation}
\begin{equation}
\frac{d\tau}{dt}=-1
\end{equation}
provided that $(4\pi\tau)^{-\frac{n}{2}}\int _{\cal{M}}e^{-f}dV=1$. Then the variation of $\cal{W}$ along the flow is

\begin{equation}
\frac{d}{dt}{\cal{W}}(g_{\mu\nu},f,\tau)=\int _{\cal{M}}\left|R_{\mu\nu}+\nabla _{\mu}\nabla _{\nu}f-\frac{1}{2\tau}g_{\mu\nu}\right|^2 e^{-f}dV
\end{equation}
We observe that $\cal{W}$ is increasing along the flow unless we are on a gradient shrinking soliton where $\cal{W}$ is constant in time along the flow. Let us assume that the manifold $\cal{M}$ is compact without boundary and set $u=\frac{1}{(4\pi\tau)^{\frac{n}{4}}}e^{-\frac{f}{2}}$ and $\int _{\cal{M}} u^{2}dV=1$. Integrating by parts the functional $\cal{W}$, we obtain

\begin{equation}
{\cal{W}} \left(g,u\right)=\int _{\cal{M}} u\left[\tau(R-4\Delta)-2\log u-n\right]u dV
\end{equation}
We note that the operator ${\cal{B}}=\tau(R-4\Delta)-2\log u-n$ acts on ``wavefunctions" which again must be strictly positive. As before, it can be shown that the lowest one is unique and it corresponds to the value of this functional.
\par
A gradient shrinking soliton, which renders $\cal{W}$ constant in time, satisfies $R_{\mu\nu}+\nabla _{\mu}\nabla _{\nu}f-\frac{1}{2\tau}g_{\mu\nu}=0$. Although there is no demand stemming from dynamics (such as the vanishing of a $\beta$-function) which would render the physical application clear, it is worth mentioning the fact that Ricci flat manifolds which serve as backgrounds or compactification spaces in string theory, can be \emph{endowed} with the structure of a shrinker with an appropriate choice of the function $f$. The easiest example is that of ${\mathbb R}^{n}$. Choosing $f=\frac{x^2}{4\tau}$ and $g_{\mu\nu}=\delta _{\mu\nu}$, we see that the condition for having a shrinker is satisfied. Finally, we can easily compute the value of $\cal{F}$ for such a soliton. For the case of a \emph{closed} manifold we have equation $R+\triangle _{\cal{M}}f-\frac{n}{2\tau}=0$ given that $(4\pi\tau)^{-\frac{n}{2}}\int _{\cal{M}}e^{-f}dV=1$. Redefining $f$ as

\begin{equation}
f\longrightarrow f'=f-\frac{n}{2\tau}\log 4\pi\tau
\end{equation}
so that $\int _{\cal{M}}e^{-f'}dV=1$, we obtain

\begin{equation}
{\cal{F}}(g,f')=\int _{\cal{M}} \left(-\triangle _{\cal{M}}f'+|\nabla f'|^{2}+\frac{n}{2\tau}\right)e^{-f'} dV=\frac{n}{2\tau}
\end{equation}
\par
As a final remark, let us comment on the relevance of the functional $\cal{F}$ in string theory (see $\cite{a_5}$ and references therein). In this context, there exists a $\sigma$-model which describes the mapping of the $2$-dim string worldsheet to a target space of dimension $D$. Conformal invariance dictates that the trace of the energy momentum tensor $T_{\alpha}^{\alpha}$ must be zero to all orders in $\alpha'$. As is well known, there is an action, called the \textit{effective action}, whose equation of motion produces the background in which the string propagates, respecting conformal invariance. Schematically, we can write this action as $S=c_{o}+\alpha'\cal{F}$ where the constant $c_{o}$ is the central charge. As a result, we observe that the aforementioned manifolds, viewed as shrinking solitons, assume this simple off-shell contribution to their central charge, to order $\alpha'$. 

\subsection{The reduced length $\cal{\ell}$}
In this subsection, we will focus on a particular energy functional introduced by Perelman, called $\cal{L}$-functional as well as its extremum, whose corresponding length is defined as the \emph{reduced length} $\ell$. Its main property is that quantity $V=\frac{1}{(4\pi\tau)^{\frac{n}{2}}}\int _{\cal{M}} e^{-\ell}$, which is called the reduced volume is monotonic along the Ricci flow. Furthermore, the introduction of this quantity can be motivated by considering a space-time description of the Ricci flow, as Perelman did, with the role of time played by $\tau$. Then $\ell$ represents some sort of ``distance" in that space-time $\cite{perel_1}$. By construction, it coincides with the shrinker potential (modulo some additive constant) in case our evolving manifold is indeed a shrinker. We will see that the soliton potential $f$ constitutes a lower bound of the $\cal{L}$-length.
\par
The easiest and most rich in properties case to consider is the two-dimensional cigar soliton, which throughout our discussion, we use as our main testing ground. Furthermore, on this background, conformally invariant (D-brane) solutions as quantum boundary states in the Hilbert space, have been extensively studied and therefore it is a good choice to delve into (see $\cite{a_8}$ and references therein). 
\par
Here we will only attempt to verify the property that the soliton potential $f$ is a lower bound for $\ell$. The full relevance of the notions surrounding the reduced length in the context of $\sigma$-models will be left for future investigations. Following the setup of $\cite{a_9}$, let us start by writing down explicitly the cigar solution of Ricci flow equation along with the evolution of the function $f$ as follows

\begin{equation}\label{1}
    g(x,y,\tau)=\frac{dx^2+dy^2}{e^{-4\tau}+x^2+y^2}
\end{equation}
\begin{equation}
    e^{-f}=1+(x^2+y^2)e^{4\tau}
\end{equation}
The Ricci scalar curvature reads

\begin{equation}
    R(x,y,\tau)=\frac{4}{1+e^{4\tau}(x^2+y^2)}
\end{equation}
Now, fix a point on the cigar. We easily observe that as $t\rightarrow+\infty$ (or $\tau\rightarrow-\infty$), then $R=4$. This means that the curvature increases as time goes on at any given point until it reaches this maximal value. At the tip $x=y=0$, it is intuitively clear that the curvature will remain at the maximal value throughout the flow. This justifies the fact that the cigar is ``burning".

\par
We proceed by computing the $\cal{L}$-length on the cigar. It is defined as

\begin{equation}
    {\cal{L}}\doteq\int _{0}^{\overline{\tau}} \sqrt{\tau}\left(R+\left|\frac{d\gamma}{d\tau}\right|^{2}_{g(\tau)}\right)d\tau
\end{equation}
where $\gamma(\tau)$ represents a curve on the cigar parameterized by $\tau$. Let us change variables $r^2=x^2+y^2$ and $\tan\theta=\frac{y}{x}$. For simplicity, by considering a radial path ($\frac{d\theta}{d\tau}=0$, $r(0)=0$) and substituting the above expressions for the Ricci scalar and the metric, we obtain

\begin{equation}
    {\cal{L}}=\int _{0}^{\overline{\tau}} \sqrt{\tau}\left[\frac{4}{1+e^{4\tau}r^2}+\frac{1}{r^2+e^{-4\tau}}\left(\frac{dr}{d\tau}\right)^2\right]d\tau
\end{equation}
Setting $re^{2\tau}=\sinh\rho $, the integral becomes

\begin{equation}
    {\cal{L}}=\int _{0}^{\overline{\tau}}\sqrt{\tau}\left[\frac{4}{\cosh^{2}\rho}+\left(\frac{d\rho}{d\tau}-2\tanh\rho\right)^2\right]d\tau
\end{equation}
If we define $\sigma\doteq 2\sqrt{\tau}$, we get

\begin{equation}
    {\cal{L}}=\int _{0}^{\overline{\sigma}}\left[\frac{\sigma^2}{\cosh^{2}\rho}+\left(\frac{d\rho}{d\sigma}-\sigma\tanh\rho\right)^2\right]d\sigma=\int _{0}^{\overline{\sigma}}\left[\sigma^2+\left(\frac{d\rho}{d\sigma}\right)^2-2\sigma\tanh\rho\frac{d\rho}{d\sigma}\right]d\sigma
\end{equation}
Given that the first two terms in the integrand are positive quantities, we get the inequality

\begin{equation}
    {\cal{L}}>\int _{0}^{\overline{\tau}}(-2\sigma\dot{\rho}\tanh\rho)d\tau=\sigma f-\int _{0}^{\overline{\sigma}}fd\sigma
\end{equation}
where $f=-2\log\cosh\rho$. The last integral is itself a negative quantity and therefore we get the final inequality

\begin{equation}
    f< \frac{1}{\sigma}{\cal{L}}
\end{equation}
The above inequality is valid for \emph{any} radial path. Therefore it would be also true for the path which renders the ${\cal{L}}$-length minimal. By definition $\ell\doteq\inf\frac{1}{\sigma}{\cal{L}}$, so we finally have $f<\ell$. Conclusively, in this computation we verify the fact that for the gradient steady soliton at hand, namely the cigar soliton, its reduced length $\ell$ is bounded from below by its potential $f$.

\section{A monotonic quantity for the mean curvature flow}
\setcounter{equation}{0}
In this section, we are considering the case of the evolution of a submanifold by its mean curvature $K$ inside a fixed ambient space. Let $S\subset\cal{M}$ be an $s$-dimensional submanifold, $\phi$ stand for its embedding and $\hat{n}$ stand for the inward normal. The mean curvature flow (MCF), assumes the form

\begin{equation}
\frac{\partial\phi}{\partial t}=H\hat{n}
\end{equation}
The fixed points of this flow are submanifolds of zero mean curvature, also known as \emph{minimal submanifolds}. As in the case of Ricci flow, a monotonic quantity and the notion of a soliton solution exist in MCF as well. Our primary focus will be the way of obtaining such a quantity from solutions of the heat equation in the ambient manifold. In physics, this equation arises as the $\beta$-function in 1-loop approximation in $\alpha '$ of an embedded brane in some given ambient space. This result was first obtained in $\cite{a_4}$. 
\par
The purpose here is to review the proof of the monotonicity formula worked out by Hamilton $\cite{a_10}$, extending an earlier result of Huisken $\cite{a_11}$, as it will serve as the basis for the generalization to Ricci solitons which will take place in the following section. Let $k$ be the solution of the backward heat equation

\begin{equation}
\frac{\partial k}{\partial t}=-\triangle _{\cal{M}}k
\end{equation}
where $k$ is normalized to 1 $\left(\int _{{\cal{M}}}k=1\right)$. We will see later that the normalization condition is of grave importance later. Let us now assume that we have a submanifold $S$, $dim S=s$ evolving by mean curvature flow in an ambient space ${\cal{M}}$, $dim {\cal{M}}=m$. We denote the normal indices  $\alpha, \beta, \gamma ...$ and tangent ones $i, j, k ... $. Then the Laplacian on $S$ is given in terms of Laplacian on ${\cal{M}}$ by the relation

\begin{equation}
\triangle _{S}k=\triangle _{\cal{M}}k-g^{\alpha\beta}\nabla _{\alpha}\nabla _{\beta}k+H^{\alpha}\nabla _{\alpha}k
\end{equation}
Now we compute

\begin{equation}
\frac{d}{dt}\int _{S}k=\int _{S} \left(\frac{\partial k}{\partial t}+H^{\alpha}\nabla _{\alpha}k-|H|^{2}k\right)=
\int _{S} \left(-\triangle _{M}k+H^{\alpha}\nabla _{\alpha}k-|H|^{2}k\right)
\end{equation}
Using (2) and  applying Stokes theorem ($\int _{S}\triangle _{S}k=0$ for a closed submanifold) we obtain

\begin{equation}
\frac{d}{dt}\int _{S}k=\int _{S}\left(-g^{\alpha\beta}\nabla _{\alpha}\nabla _{\beta}k+2H^{\alpha}\nabla _{\alpha}k-| H|^{2}k\right)
\end{equation}
Adding and subtracting the quantity $\frac{\nabla _{\alpha}k\nabla ^{\alpha}k}{k}$ we get

\begin{equation}
\frac{d}{dt}\int _{S}k=\int _{S}-\left(|H|^{2}k-2H^{\alpha}\nabla _{\alpha}k+\frac{\nabla _{\alpha}k \nabla ^{\alpha}k}{k}\right)+
\int _{S}\left(\frac{\nabla _{\alpha}k \nabla ^{\alpha}k}{k}-\nabla _{\alpha}\nabla ^{\alpha}k\right)
\end{equation}
This becomes

\begin{equation}
\frac{d}{dt}\int _{S}k=\int _{S}-\left|H-\frac{\nabla^{\bot} k}{k}\right|^{2}k+\int _{S}\left(\frac{\nabla _{\alpha}k \nabla ^{\alpha}k}{k}-\nabla _{\alpha}\nabla _{\alpha}k\right)
\end{equation}
Finally one obtains

\begin{equation}
 \begin{split}
     &\frac{d}{dt}\tau^{\frac{m-s}{2}}\int _{S}k+\tau^{\frac{m-s}{2}}\int _{S}\left|H-\frac{\nabla^{\bot} k}{k}\right|^{2}k
	  \\
    &\qquad +\tau^{\frac{m-s}{2}}\int _{S}g^{\alpha\beta}\left(\nabla _{\alpha}\nabla _{\beta}k-\frac{\nabla _{\alpha}k \nabla _{\beta}k}{k}+\frac{1}{2\tau}g_{\alpha\beta}k\right)=0
   \end{split}
\end{equation}
We can make the substitution $f=-\log k$ and rewrite the final result as

\begin{equation}
 \begin{split}
     &\frac{d}{dt}\tau^{\frac{m-s}{2}}\int _{S}e^{-f}+\tau^{\frac{m-s}{2}}\int _{S}\left|H+\nabla^{\bot} f\right|^{2}e^{-f}
	  \\
    &\qquad +\tau^{\frac{m-s}{2}}\int _{S}g^{\alpha\beta}\left(-\nabla _{\alpha}\nabla _{\beta}f+\frac{1}{2\tau}g_{\alpha\beta}\right)e^{-f}=0
   \end{split}
\end{equation}
At this point we invoke the Hamilton's matrix Harnack inequality according to which the last integrand is non-negative under the assumption that the manifold $\cal{M}$ is Ricci parallel ($\nabla _{\mu}R_{\nu\rho}=0$) and has weakly positive sectional curvature ($R_{\mu\nu\rho\lambda}V^{\mu}W^{\nu}V^{\rho}W^{\lambda}\geq 0$ where $V^{\mu},W^{\nu}$ are two arbitrary vectors spanning a plane) $\cite{a_10}$. Therefore we obtain

\begin{equation}
\frac{d}{dt}\tau^{\frac{m-s}{2}}\int _{S}e^{-f}\leq 0
\end{equation}
It is worth mentioning that in the case of zero sectional curvature, namely ${\cal{M}}\equiv{\mathbb R}^{n}$, the fundamental solution of the heat kernel equation saturates the Harnack inequality and had been obtained by Huisken $\cite{a_11}$.

\section{Ambient space moving by Ricci flow}
\subsection{Derivation of the monotonic quantity}
We now proceed to examine the case where the metric of the ambient space evolves by Ricci flow and the embedded space evolves by MCF. We will try to construct a monotonic quantity along this combined flow in the same spirit as before. The strategy will be to integrate a solution of the heat equation of the evolving ambient space over an \emph{also} evolving embedded space and see what modifications we will be confronted with. The backward heat equation is modified to

\begin{equation}
\frac{\partial k}{\partial t}=-\triangle _{\cal{M}}k+Rk
\end{equation}
where by $R$ we denote the scalar Ricci curvature of the ambient manifold and again $\int _{{\cal{M}}}k=1$. Therefore

\begin{equation}
\frac{d}{dt}\int _{S}k=\int _{S} \frac{\partial k}{\partial t}+H^{\alpha}\nabla _{\alpha}k-|H|^{2}k-R^{\top}k
\end{equation}
where $R^{\top}$ is the induced Ricci scalar curvature (to be more precise, $R^{\top}=\sum {R(V,V)}, V\in{T_{p}S}$, where $T_{p}S$ is the tangent bundle of $S$ and the summation is over all V). Again, we substitute the heat kernel equation, we use the Stokes theorem and noting that $R=R^{\top}+R^{\bot}$ where $R^{\bot}$ is the Ricci scalar computed in the normal bundle (meaning $R^{\bot}=\sum R(W,W), W\in{N_{p}S}$, where $N_{p}S$ is the normal bundle and the summation is over all W), we obtain

\begin{equation}
\frac{d}{dt}\int _{S}k=\int _{S}-g^{\alpha\beta}\nabla _{\alpha}\nabla _{\beta}k+2H^{\alpha}\nabla _{\alpha}k-|H|^{2}k+R^{\bot}k
\end{equation}
Finally we get

\begin{equation}
 \begin{split}
     &\frac{d}{dt}\tau^{\frac{m-s}{2}}\int _{S}k+\tau^{\frac{m-s}{2}}\int _{S}\left|H-\frac{\nabla^{\bot} k}{k}\right|^{2}k
	  \\
    &\qquad +\tau^{\frac{m-s}{2}}\int _{S}g^{\alpha\beta}\left(\nabla _{\alpha}\nabla _{\beta}k-\frac{\nabla _{\alpha}k \nabla _{\beta}k}{k}+\frac{1}{2\tau}g_{\alpha\beta}k-R_{\alpha\beta}k\right)=0
   \end{split}
\end{equation}
We can rewrite this by setting $-f=\log k$ as

\begin{equation}
 \begin{split}
     &\frac{d}{dt}\tau^{\frac{m-s}{2}}\int _{S}e^{-f}+\tau^{\frac{m-s}{2}}\int _{S}\left|H+\nabla^{\bot} f\right|^{2}e^{-f}
	  \\
    &\qquad +\tau^{\frac{m-s}{2}}\int _{S}g^{\alpha\beta}e^{-f}\left(-\nabla _{\alpha}\nabla _{\beta}f+\frac{1}{2\tau}g_{\alpha\beta}-R_{\alpha\beta}\right)=0
   \end{split}
\end{equation}
From the above result we observe that if the ambient manifold is a shrinker satisfying $R_{\mu\nu}+\nabla _{\mu}\nabla _{\nu}f-\frac{1}{2\tau}g_{\mu\nu}=0$, then $\nabla _{\alpha}\nabla _{\beta}f-\frac{1}{2\tau}g_{\alpha\beta}+R_{\alpha\beta}=0$. Therefore in this case the monotonic quantity is indeed the solution of the conjugate heat equation in the ambient space integrated over a submanifold evolving by MCF. There is a similar expression for the case of a steady soliton. Instead of the above equation, we could have gotten

\begin{equation}
 \begin{split}
     &\frac{d}{dt}\int _{S}e^{-f}+\int _{S}\left|H+\nabla^{\bot}\log f\right|^{2}e^{-f}
	  \\
    &\qquad +\int _{S}g^{\alpha\beta}e^{-f}(-\nabla _{\alpha}\nabla _{\beta}f-R_{\alpha\beta})=0
   \end{split}
\end{equation}
Therefore it is obvious that if the ambient manifold is a steady soliton (meaning $R_{\mu\nu}+\nabla _{\mu}\nabla _{\nu}f=0$), then by similar reasoning as above, we get a monotonic quantity along the combined flow. It is also important to note that the soliton property is preserved. This can be seen by taking the trace of the soliton equations and substitute them into the heat equation (for the shinker we should use $f\rightarrow f'=f+\frac{n}{2}\log 4\pi\tau$ to keep up with the standard notation). We will end up with an equation of the form $\frac{\partial f}{\partial t}=|\nabla f|^{2}$ which is the standard evolution of the potential of a gradient soliton.

\subsection{Examples}
To get a more tangible feel of the above considerations, we consider simple examples of arbitrary closed curves supported on Ricci solitons.

\subsubsection{Curves in Euclidean space}
The easiest to start with is that of ${\mathbb R}^{2}$ as our ambient manifold (which is shrinker by choosing $g=\frac{r^2}{4\tau}$ and $r^2=x^2+y^2$) and an embedded sphere ${\mathbb S}^{1}$ evolving by MCF. We can easily check that, solving the MCF equation, we get $r^2=2\tau$, with $\tau=0$ being the ``crunch" time. It is easily verified that choosing $g=f$, $H+\nabla^{\bot} f=0$ holds true. This is a result which had been obtained by Huisken $\cite{a_11}$. As a consequence this shrinking curve is indeed a soliton of MCF and satisfies

\begin{equation}
\frac{d}{dt}\sqrt{\tau}\int _{S}e^{-f}=0
\end{equation}
\par
The generalization to higher dimensional Euclidean space is straightforward. In this case a soliton of MCF would satisfy

\begin{equation}
\frac{d}{dt}\tau^{\frac{m-s}{2}}\int _{S}e^{-f}=0
\end{equation}

\subsubsection{Curves inside the sphere}
Let us consider the case of the sphere ${\mathbb S}^2$ of radius $r(\tau)$. The solution of the Ricci flow yields immediately $r^2=2\tau$. From the general equation satisfied by a shrinker we see that the potential function $f$ is \emph{at most} a function of $\tau$. From the normalization condition $\int _{{\cal{M}}}k=1$ we obtain $k=\frac{1}{8\pi\tau}$ so $f=\log 8\pi\tau$. From the results obtained previously, we get

\begin{equation}
\frac{d}{dt}\frac{1}{8\pi\sqrt{\tau}}\int _{S}ds=-\frac{1}{8\pi\sqrt{\tau}}\int _{S}\left|H\right|^{2}ds
\end{equation}
where $ds$ is the line element of the curve.
\par
In the case of an evolving submanifold of codimension  $c$ inside ${\mathbb S}^{n}$, we have that $f=\log V_{{\mathbb S}^n}(2\tau)^{\frac{n}{2}}$, and the above formula generalizes to

\begin{equation}
\frac{d}{dt}\frac{1}{2^{\frac{n}{2}}V_{{\mathbb S}^n}\tau^{\frac{c}{2}}}\int _{S}da=-\frac{1}{2^{\frac{n}{2}}V_{{\mathbb S}^n}\tau^{\frac{c}{2}}}\int _{S}|H|^{2}da
\end{equation}
where $da$ is the volume of the submanifold and $V_{{\mathbb S}^n}$ is the volume of ${\mathbb S}^n$. Therefore we see that this solution of the conjugate heat kernel equation, integrated over an evolving submanifold, stays monotonic along the combined flow. Note that the submanifold is not a soliton of MCF.

\subsubsection{Curves inside the cigar}
What about the two dimensional cigar soliton? In $\cite{a_12}$, one closed shrinking solution has been found. It assumes the form $\cosh\rho=e^{\tau}$ where $\rho$ is the distance from the tip of the cigar. It has the form of  circle placed perpendicularly to the axis of symmetry of the cigar. We observe $\tau=0$ is the ``crunch" time where the circle disappears at the tip where $\rho=0$. Now, let us rewrite our solution as

\begin{equation}\label{2}
\cosh\rho=e^{\tau}\Rightarrow\sinh^2\rho=e^{2\tau}-1\Rightarrow r^2e^{4\tau}=e^{2\tau}-1\Rightarrow \left(x^2+y^2\right)e^{4\tau}=e^{2\tau}-1
\end{equation}
(see change of variables in section 2.2). We also define $h(\tau)=e^{-2\tau}(1-e^{-2\tau})$ so that $x^2+y^2=h(\tau)$. Therefore for the soliton potential of the cigar we obtain,

\begin{equation}
e^{-f}=1+(x^2+y^2)e^{4\tau}=e^{2\tau}
\end{equation}
Now the time derivative of the integral of the above quantity over the submanifold, which is the contracting circle on the cigar, yields the result

\begin{equation}
\frac{d}{dt}\int _{S}e^{-f}=\frac{d}{dt}\rho^2(\tau)\int _{0}^{2\pi}e^{2\tau}=-4\pi e^{2\tau}(\rho^2(\tau)+\rho\frac{d\rho}{d\tau})<0
\end{equation}
where $\rho$ is the distance from the tip of the cigar. The first term in the parenthesis is clearly positive and the second is positive as well since the distance increases from the tip in ``inverse" time. This renders the functional under consideration monotonic.
\par

Can this solution be obtained from the demand $H+\nabla^{\bot} f=0$? Let us compute the mean curvature vector for this solution. The general formula for an arbitrary curve embedded in a two dimensional surface is

\begin{equation}
H=\frac{{\sqrt{\Omega}\phi''}}{\sqrt{\left(1+\phi'^2\right)^3}}+\frac{1}{2\sqrt{\Omega}\sqrt{\left(1+\phi'^2\right)}}\left(\partial _{y}\Omega-\phi'\partial _{y}\Omega\right)
\end{equation}
In this expression we have assumed that the metric of the ambient space has the form

\begin{equation}
ds^2=\frac{1}{\Omega}\left(dx^2+dy^2\right)
\end{equation}
and the curve is given as $y=\phi(x)$. Applying this formalism to our case, we observe that $\Omega$ obtained form the metric~\eqref{1} simplifies to $\Omega=e^{-2\tau}$. The equation~\eqref{2} can be solved for $y$ and then the expression for curvature can be used. The result is

\begin{equation}
H=-\frac{\sqrt{\Omega}}{\sqrt{h(\tau)}}+\frac{\sqrt{h(\tau)}}{\sqrt{\Omega}}
\end{equation}
Substituting the expressions for $\Omega$ and $h(\tau)$, we obtain

\begin{equation}
H=-\frac{1}{\sqrt{1-e^{-2\tau}}}+\sqrt{1-e^{-2\tau}}
\end{equation}
We note that the mean curvature $H$ depends only on time since every point on the circle has the same extrinsic curvature due to symmetry. Let us check two limits of our solution. First, we take $\tau\rightarrow\infty$, and find $H=0$ as expected, since we have a circle wrapped on cylinder and placed perpendicularly with respect to its axis of symmetry. Also, when we take $\tau\rightarrow 0$, we get $H\rightarrow-\infty$. This means that the circle has shrunk to a point at the tip of the cigar.
\par

Returning to our original question, let us make the choice $f=\ln\tanh\rho$. We compute $\nabla^{\bot} f=\partial _{\rho}f=\coth\rho-\tanh\rho$. But using~\eqref{2}, $\coth\rho=\frac{e^{\tau}}{\sqrt{e^{2\tau}-1}}$ and $\tanh\rho=\frac{\sqrt{e^{2\tau}-1}}{e^{\tau}}$. Therefore we found a particular $f$ such that $H+\nabla^{\bot} f=0$. Does this mean that the new ``dilaton" we constructed leads to a monotonic quantity along the combined flow? The answer is no. This is because, even though we had been able to come up with an appropriate reparameterization which ``freezes" the MCF for the solution at hand, this does not mean that we found a new solution of backward conjugate heat equation of the ambient manifold. Therefore the conclusion is that only through solutions of the backward conjugate heat equation can one obtain a monotonic quantity.

\section{Discussion}
In this work we have examined several aspects of monotonic functionals along the combined Ricci and the mean curvature flows. Our primary focus has been the solutions of the \emph{backward} conjugate heat equation in some evolving space integrated over some evolving embedded space. We have seen that for the case of a Ricci soliton, a normalized solution of the conjugate heat equation integrated over some evolving submanifold by MCF stays monotonic along the combined flow. 
\par
It would be useful to comment on the physical relevance of the previous considerations. As we have seen above, in the case where the ambient manifold is a Ricci soliton, the final result for the shrinker and the steady one is

\begin{equation}
\frac{d}{dt}\tau^{\frac{m-s}{2}}\int _{S}e^{-f}=-\tau^{\frac{m-s}{2}}\int _{S}\left|H+\nabla^{\bot} f\right|^{2}e^{-f}
\end{equation}
\begin{equation}
\frac{d}{dt}\int _{S}e^{-f}=-\int _{S}\left|H+\nabla^{\bot}f\right|^{2}e^{-f}
\end{equation}
respectively. When do these quantities remain constant under the time evolution? They do remain constant when $H+\nabla^{\bot} f=0$ (this is a vector equation). It is obvious that this is a fixed point of the MCF \emph{modulo} a  diffeomorphism generated by $f$. We recognize this functional as being of the same form as the Dirac-Born-Infeld action (DBI) in the absence of the gauge field $F_{\mu\nu}$ and the antisymmetric one $B_{\mu\nu}$. The difference is that the evolution of the dilaton in string theory (dilaton $\beta$-function) has not the same evolution as the potential function $f$ which has been used here.
\par
We note the analogy which exists between the role of the DBI action for the evolution of a brane and the functional ${\cal{F}}$ we encountered in the previous section. The Ricci flow and the MCF, both modulo diffeomorphisms, are gradient flows of ${\cal{F}}$ and the DBI action respectively. The most important property that they share is that of monotonicity. While the functionals introduced by Perelman stay monotonic under \emph{any} Ricci flow, the DBI action remains monotonic along the combined Ricci flow and MCF provided that the ambient space is a Ricci soliton. It is also worth mentioning that the above construction serves as a step toward generalizing the $c$-theorem $\cite{zamo_1}$ in the presence of defects (branes). 
\par
More generally, the challenge is to find an "entropy" functional for the case of an ambient space which is not a Ricci soliton. What happens in then? To the best of our knowledge, the answer is not known. For a general Ricci flow accompanied with the backward conjugate heat equation, the existence of the Harnack inequality is currently under investigation $\cite{a_14}$. Recently Ni $\cite{a_13}$ proved that for the case of Kahler-Ricci flow the Harnack inequality holds, namely $\partial\overline{\partial}\log k-Ric+\frac{1}{2\tau}g\geq 0$. Yet, this inequality holds under the assumption that $k$ is a solution of the \emph{forward} conjugate heat equation for the case of the Kahler-Ricci flow. This would cover the case where we have a compact closed two-dimensional background such as the Sausage model (Rosenau solution in the Mathematical literature) on which a closed curve is supported. However, the combination of Ricci flow in the bulk, (MCF) for the evolution of the submanifold and the forward heat equation requires a study in its own right and will be attempted in the future.

\vskip 0.4in
\centerline{\sf{Acknowledgments}}
This work was supported in part by the bilateral research grant ``Gravity, Gauge Theories and String Theory" (06FR-020) for the Greek-French scientific cooperation. I would like to thank the organizers of the trimester on ``Ricci curvature and Ricci flow" at the Poincare Institute in Paris as well as C. Sourdis for discussions. Also, I would like to thank Carlo Mantegazza for numerous enlightening discussions as well as for the warm hospitality in Scuola Normale Superiore in Pisa. Most of all, I express my deepest gratitude to my adviser Ioannis Bakas for constant stimulation and for providing an environment conducive to research.
\newpage
\section*{Appendix A}
In this appendix we compute the evolution of the induced volume element $d\alpha$ under Ricci and mean curvature flow. We start with the induced metric on the submanifold

\begin{equation}
g_{ij}=\frac{\partial X^{\mu}}{\partial Y^{i}}\frac{\partial X^{\nu}}{\partial Y^{j}}g_{\mu\nu}
\end{equation}
where $X^{\mu}$ describes the embedding. Taking the time derivative we obtain

\begin{equation}
\frac{dg_{ij}}{dt}=2\frac{\partial}{\partial Y^{i}}\frac{dX^{\mu}}{dt}\frac{\partial X^{\nu}}{\partial Y^{j}}g_{\mu\nu}+\frac{\partial X^{\mu}}{\partial Y^{i}}\frac{\partial X^{\nu}}{\partial Y^{j}}\frac{dg_{\mu\nu}}{dt}
\end{equation}
Now we use the following facts:

\begin{equation}
  \begin{split}
\frac{dg_{\mu\nu}}{dt}=\frac{\partial g_{\mu\nu}}{\partial t}+\frac{\partial g_{\mu\nu}}{\partial X^{\lambda}}\frac{d X^{\lambda}}{dt} \\
\frac{\partial g_{\mu\nu}}{\partial X^{\lambda}}=2 g_{\mu\rho}\Gamma ^{\rho}_{\lambda\nu} \\
\frac{\partial g_{\mu\nu}}{\partial t}=-2R_{\mu\nu} \\
\frac{\partial X^{\mu}}{\partial t}=H^{\sigma}\eta _{\sigma}^{\mu}
   \end{split}
\end{equation}
where in the last equation $\eta_{\sigma}^{\mu}$ is the unit inward normal. The subscript $\sigma$ parametrizes the normal directions. Now the time derivative of the induced metric becomes

\begin{equation}
\frac{dg_{ij}}{dt}=-2 g_{\alpha\beta}H^{\alpha}K^{\beta}_{ij}-2 R_{ij}
\end{equation}
where $K^{\beta}_{ij}$ is the second fundamental form. Finally, the time derivative of the induced volume element $d\alpha$ becomes

\begin{equation}
\frac{d}{dt}d\alpha=\frac{1}{2} g^{ij}\frac{dg_{ij}}{dt}d\alpha=(-|H|^{2}-R^{\top})d\alpha
\end{equation}

\end{document}